\documentclass{LT23auth}
\usepackage{graphicx}

\def\vecS{\mbox{\boldmath{S}}}
\begin{document}

\begin{frontmatter}

\title{Spin-wave Hamiltonian in double-exchange systems}

\author[address1]{Nobuo Furukawa\thanksref{thank1}},
\author[address2]{Yukitoshi Motome}

\address[address1]{Department of Physics, Aoyama Gakuin University, 
Setagaya, Tokyo 157-8572, Japan}

\address[address2]{Tokura Spin SuperStructure Project, ERATO,
Japan Science and Technology Corporation,
c/o National Institute of Advanced Industrial Science and Technology,
Tsukuba, Ibaraki 305-8562, Japan}

\thanks[thank1]{Corresponding author. E-mail: furukawa@phys.aoyama.ac.jp}

\begin{abstract}
A simple derivation of the effective spin-wave
Hamiltonian for a double-exchange system with
infinitely large Hund's-rule coupling is demonstrated.
The formalism can be applied to models with arbitrary range of hopping
as well as those with randomness.
The result shows that, within the leading order of the $1/S$ expansion,
one magnon excitation spectrum can be 
described by the Heisenberg model.
\end{abstract}

\end{frontmatter}

Double-Exchange (DE) model 
\cite{Zener1951,Anderson1955}
has been introduced to study the metallic ferromagnetism 
in perovskite manganese oxides.
One-magnon excitation 
spectrum has been investigated using the
spin-wave approximation \cite{Furukawa1996,Golosov2000,Shannon2001}.
Recently, spin excitation spectrum of
the DE systems with randomness
has been studied using the Green's-function formalism
\cite{Motome2002}.

In this paper, we demonstrate a simple derivation
of the effective spin-wave Hamiltonian for the DE
systems, which can be applied to generic cases with
arbitrary hopping range or with randomness.
Within the leading order of the $1/S$ expansion, the results
are equivalent to those by the Green's function formalism.
The advantage of this derivation is that it provides an
intuitive understanding of the ferromagnetic exchange coupling
mediated by the DE interactions, compared to the
Green's-function formalism.

We begin with the DE model in the limit of large Hund's-rule coupling
with localized spins being treated as classical spins.
For the moment, we consider a system without randomness.
In this limit, local spin quantization axes for
conduction electrons are taken parallel
to the localized spin in each site, and electrons
with antiparallel spin states  to localized spins are projected out. Then,
the transfer integral between sites $i$ and $j$
depends on the relative angle of localized spins at corresponding sites
$\vecS_i$ and $\vecS_j$  as \cite{Anderson1955}
\begin{eqnarray}
&&t_{ij}(\vecS_i,\vecS_j) = 
\nonumber \\
&&\quad  t_{ij}{}^0 \left\{
  \cos\frac{\theta_i}2 \cos\frac{\theta_j}2
   + \sin\frac{\theta_i}2 \sin\frac{\theta_j}2 
  e^{- {\rm i} (\phi_i-\phi_j)}\right\}.
\end{eqnarray}
Here $\theta$ and $\phi$ are defined by the direction of
the localized spin $\mbox{\boldmath{S}}$ as
\begin{eqnarray}
  S_i^x &=& S \sin \theta_i \cos \phi_i, \nonumber \\
  S_i^y &=& S \sin \theta_i \sin \phi_i, \\
  S_i^z &=& S \cos \theta_i,\nonumber 
\end{eqnarray}
while $t_{ij}{}^0=t_{ji}{}^0$ is the transfer integral between
sites $i$ and $j$ in the absence of the
DE interaction.
The Hamiltonian is given by
\begin{equation}
 H = -\sum_{ij}
  \left[ t_{ij}(S_i,S_j) c_i^\dagger c_j + t_{ji}(S_j,S_i) c_j^\dagger c_i \right].
\end{equation}
Note that we explicitly treat the complex transfer integral
in this formalism.

The absolute value of the transfer integral is rewritten as
\begin{eqnarray}
  \frac{\left|t_{ij}(\vecS_i,\vecS_j)\right|}{t_{ij}{}^0} 
 &=& \sqrt{
      \frac12 + \frac1{2S^2} \mbox{\boldmath{S}}_i \cdot \mbox{\boldmath{S}}_j
     }.
  \label{Abstij}
\end{eqnarray}
On the other hand, the imaginary part becomes as
\begin{eqnarray}
 \frac{ \mbox{\sl Im\,} t_{ij}}{t_{ij}{}^0}
 &=& \frac12 \sqrt{ 
    \frac{S^2}{(S+S_i^z)(S+S_j^z)}
   } \, \frac{S_i^y S_j^x - S_i^x S_j^y}{S^2}.
  \label{Imtij}
\end{eqnarray}
Note that $\mbox{\sl Im\,}t_{ij}$ is antisymmetric with respect to
the exchange of $i$ and $j$,
as expected from $t_{ij} = t_{ij}^*$.

Now, we apply the spin-wave approximation to this Hamiltonian.
We consider the spin-wave excitation from
the perfectly-polarized ferromagnetic ground state
where the spins align along the $z$ direction.
Namely, we replace the spin variables $\vecS_i$ by
the Holstein-Primakoff transformation
\begin{eqnarray}
S_i^x &\simeq& \sqrt{\frac S2}\left( a_i^\dagger + a_i \right),
 \nonumber \\
  \label{def1/S}
S_i^y &\simeq& {\rm i}\sqrt{\frac S2}\left( a_i^\dagger -a_i \right),
 \\
S_i^z &=& S - a_i^\dagger a_i, \nonumber
\end{eqnarray}
and take account of the expansion by $1/S$ up to the leading order $O(1/S)$.
By substituting Eqs.~(\ref{def1/S}) into Eq.~(\ref{Abstij}), we obtain
\begin{equation}
  \frac{|t_{ij}|}{t_{ij}{}^0} = 1+ 
   \frac1{4S}(a_i^\dagger a_j + a_j^\dagger a_i 
              - a_i^\dagger a_i - a_j^\dagger a_j) + O(\frac1{S^2}).
\end{equation}
In the same manner, we obtain the imaginary part as
\begin{equation}
 \frac{ \mbox{\sl Im\,} t_{ij}}{t_{ij}{}^0} =
 \frac{1}{4S} \left( a_i^\dagger a_j - a_j^\dagger a_i \right)
  + O(\frac1{S^2}).
\end{equation}
Since the imaginary part is $O(1/S)$,
if we denote $t_{ij} = |t_{ij}| \exp{({\rm i} \Phi_{ij})}$,
we obtain
\begin{eqnarray}
{\rm e}^{{\rm i}\Phi_{ij}} 
  &=& 1 + {\rm i} \Phi_{ij} + O(\frac1{S^2}) \nonumber \\
  &=& 1 + {\rm i} \frac{ \mbox{\sl Im\,} t_{ij}}{|t_{ij}|}
   + O(\frac1{S^2}).
\end{eqnarray}
Therefore, we obtain the magnon-electron Hamiltonian 
in the spin-wave approximation up to $O(1/S)$ as
%
\begin{eqnarray}
 H &\simeq& - \sum_{ij} t_{ij}{}^0
\left[ 1 + 
   \frac1{4S}(a_i^\dagger a_j + a_j^\dagger a_i 
 - a_i^\dagger a_i - a_j^\dagger a_j)
    \right] \nonumber\\
  && \qquad\qquad\qquad\times
 \left(c_i^\dagger c_j + c_j^\dagger c_i\right) \nonumber\\
&&
  +\frac{1}{4S} \sum_{ij}  t_{ij}{}^0
    \left(a_i^\dagger a_j - a_j^\dagger a_i \right) 
  \left(c_i^\dagger c_j -c_j^\dagger c_i \right).
\label{defHam1/S}
\end{eqnarray}


To obtain the effective Hamiltonian for magnons,
we trace out the fermion degrees of freedom in Eq.~(\ref{defHam1/S}).
Up to $O(1/S)$, 
the result is given by replacing terms $c_i^\dagger c_j$
by $\langle c_i^\dagger c_j\rangle$. Here, the
expectation value should be taken for the ferromagnetic ground state
without any magnon, whose Hamiltonian is described by
\begin{eqnarray}
 H_0 = - \sum_{ij} t_{ij}{}^0
  \left(c_i^\dagger c_j + c_j^\dagger c_i\right).
\label{defHam0}
\end{eqnarray}
In the perfectly-polarized ground state without
degeneracies, 
the relation $\langle c_i^\dagger c_j\rangle =
\langle c_j^\dagger c_i\rangle$ generally holds 
since the expectation value is real.
Then, the second term in Eq.~(\ref{defHam1/S}) vanishes and
we finally obtain the effective spin-wave Hamiltonian as
\begin{eqnarray}
 H_{\rm eff} &=& -\frac{1}{2S} \sum_{ij} t_{ij}{}^0
   \langle c_i^\dagger c_j \rangle 
   \nonumber \\
 && \times \left(
    a_i^\dagger a_j + a_j^\dagger a_i 
              - a_i^\dagger a_i - a_j^\dagger a_j
   \right),
 \label{defHamSW}
\end{eqnarray}
up to irrelevant constants.
In the uniform system with nearest neighbor hoppings
where $\langle c_i^\dagger c_j\rangle$ is constant,
this Hamiltonian gives a cosine-like dispersion as previously obtained
by the different method \cite{Furukawa1996}.

Let us discuss the relation with the Heisenberg model.
Comparing with the spin-wave approximation (\ref{def1/S})
of the Heisenberg model $  H_{\rm Heis} =
 -2 \sum J_{ij} \mbox{\boldmath{S}}_i \cdot \mbox{\boldmath{S}}_j 
$, 
we see that the magnon Hamiltonian (\ref{defHamSW}) can be
reproduced by the Heisenberg model with exchange couplings
\begin{equation}
 J_{ij} = \frac {t_{ij}{}^0}{4S^2} \langle c_i^\dagger c_j \rangle,
  \label{defJ}
\end{equation}
within the leading order of $1/S$ expansion.
Since ${t_{ij}{}^0}  \langle c_i^\dagger c_j \rangle$ describes
the local kinetic energy, 
Eq.~(\ref{defJ}) gives the relation between 
the DE ferromagnetic interaction and
the kinetics of conduction electrons. 

Finally, we note that Eq.~(\ref{defJ}) has been derived
for generic electronic hoppings $t_{ij}{}^0$,
which includes the cases with 
arbitrary hopping range or with random hopping integrals.
Let us consider the case with site-diagonal
random potential for conduction electrons.
Since this type of potential does not couple to the
spin-wave operators within the leading terms of $1/S$,
the spin-wave Hamiltonian can be
similarly obtained.
Namely, the Hamiltonian (\ref{defHam0}) should be replaced by that
with random potential, and 
expectation values  $ \langle c_i^\dagger c_j \rangle$ should be
taken by the ground state of the replaced Hamiltonian.
We also see that $J_{ij}\ne 0$ for 
Anderson-localized insulating systems where 
$ \langle c_i^\dagger c_j \rangle \ne 0$.
Namely, DE ferromagnetism can also exist  in such non-metallic systems.

%
%

\end{document}